\documentclass[a4paper,11pt]{article}
\usepackage{pos}
\usepackage{amsmath}
\usepackage{amsfonts}
\usepackage{mathrsfs}
\usepackage{bbm}
\usepackage[normalem]{ulem}

\usepackage{latexsym}
\usepackage{amsthm}
\usepackage[all,knot]{xy}
\xyoption{arc}

\hypersetup{
  hypertexnames=false,
  colorlinks=true,
  pdfauthor={Hyungrok Kim,Christian S\"amann},
  pdftitle={},
  pdfsubject={hep-th math-ph},
  breaklinks=true
}

\usepackage{tikz}
\usetikzlibrary{matrix,cd,decorations.markings}
\tikzcdset{arrow style=tikz}
\pgfmathdeclarefunction*{axis_height}{0}{\begingroup\pgfmathreturn.25em\endgroup}
\pgfmathdeclarefunction*{rule_thickness}{0}{\begingroup\pgfmathreturn.06em\endgroup}

\usepackage{mathtools}
\usepackage[all,knot]{xy}
\xyoption{arc}

\newcommand{\makecommand}[3]{%
  \foreach \i in #3 {%
    \expandafter\xdef\csname #1\i\endcsname{\noexpand#2{\unexpanded\expandafter{\i}}}%
  }%
}

\makecommand{fr}{\mathfrak}{{c,der,g,h,A,X,gl,sl,o,so,osp,u,su,sp,spin,string,Poisson,td,tb}}

\makecommand{sf}{\mathsf}{{id,String,Lie,Hom,SU,SO,Cat,Set,Spin,Pin,inv,CE,Diff,HSymp,TD,TB,pr,Aut,GL,SL,IO,Mat,Sym,Inn,Out,AUT}}

\makecommand{rm}{\mathrm}{{Ham,diag,im}}

\newcommand{\latinalphabet}{A,a,B,b,C,c,d,D,E,e,F,f,G,g,H,h,I,i,J,j,K,k,L,l,M,m,N,n,O,o,P,p,Q,q,R,r,S,s,T,t,U,u,V,v,W,w,X,x,Y,y,Z,z}
\newcommand{\caplatinalphabet}{A,B,C,D,E,F,G,H,I,J,K,L,M,N,O,P,Q,R,S,T,U,V,W,X,Y,Z}
\makecommand{I}{\mathbbm}{\latinalphabet}
\makecommand{bf}{\mathbf}{\latinalphabet}
\makecommand{bm}{\bm}{\latinalphabet}
\makecommand{ca}{\mathcal}{\caplatinalphabet}
\makecommand{fr}{\mathfrak}{\latinalphabet}
\makecommand{rm}{\mathrm}{\latinalphabet}
\makecommand{sf}{\mathsf}{\latinalphabet}
\makecommand{sc}{\mathscr}{\latinalphabet}

\newcommand{\sfGO}{\sfG\sfO}
\newcommand{\scTD}{\scT\!\scD}

\newcommand{\scGO}{\scG\!\scO}
\newcommand{\acton}{\triangleright}
\newcommand{\unit}{\mathbbm{1}}   			
\newcommand{\eand}{~~~\mbox{and}~~~}

\title{T-duality as Correspondences of Categorified Principal Bundles with Adjusted Connections}
\ShortTitle{T-duality as Correspondences of Categorified Principal Bundles}

\author[a]{Hyungrok Kim}
\author*[a]{Christian Saemann}

\affiliation[a]{Maxwell Institute and Department of Mathematics\\
  Heriot--Watt University, Edinburgh EH14 4AS, United Kingdom}

\emailAdd{hk55@hw.ac.uk}
\emailAdd{c.saemann@hw.ac.uk}

\abstract{We provide a pedagogical introduction to the theory of principal 2-bundles with adjusted connections and show how they enter the description of geometric and non-geometric T-dualities as proposed in arXiv:2204.01783. This description combines the torus fibrations as well as the gerbe containing the Kalb--Ramond $B$-field into a single geometric object, a particular case of a non-abelian gerbe. The $B$-field and the metric are encoded in the connection of this categorified principal bundle, and a T-duality is described as a particular span or correspondence of such bundles. The formalism is manifestly covariant under the full T-duality group, and it readily reproduces key examples from the literature.}

\FullConference{%
  Corfu Summer Institute 2022 "School and Workshops on Elementary Particle Physics and Gravity",\\
  28 August - 1 October, 2022\\
  Corfu, Greece\\[0.5cm]Report No: EMPG--23--05}

\tableofcontents

\begin{document}
  \maketitle

  \section{Introduction}
  
  String theory is usually formulated in terms of 2-dimensional conformal field theory sigma-models with input data the \emph{target space}, i.e.~a Riemannian manifold $(P,G)$, and the \emph{Kalb--Ramond $B$-field}, a connection on an abelian bundle gerbe $\scG$ on $P$.\footnote{In addition, one may consider the dilaton and Ramond--Ramond fields, but we will completely ignore them in our discussion.} In many cases, the target space $P$ is a torus bundle over some Riemannian manifold $X$, and the metric on $P$ is the Kaluza--Klein metric derived from a connection on $P$ a the metric on the base space $X$. \emph{T-duality}~\cite{Buscher:1987sk,Buscher:1987qj}, see also~\cite{Giveon:1994fu} is now a particular exchange of the data $(P,\scG)$ with data $(\tilde P,\tilde \scG)$ such that the resulting sigma models are physically equivalent. Under this exchange, radii of the torus fibers are inverted, and the momentum and winding modes along the torus fibers are exchanged. The latter perspective highlights the fact that T-duality is a symmetry that truly sets apart string theory from particle physics, as only strings can be wound around compact directions. 
  
  There are manifold reasons for studying T-duality, from seeking a better understanding of strings themselves to finding physical realizations of the Fourier-Mukai transform. Our primary motivation is that T-duality provides a very useful example to apply and test our notion of principal 2-bundles with adjusted connections~\cite{Saemann:2019dsl,Kim:2019owc,Borsten:2021ljb,Rist:2022hci}. A second motivation comes from the fact that higher geometry seems to be able to ``resolve'' the non-geometric string backgrounds arising in the context of T-duality.
  
  \subsection{Topological T-duality}
  
  In a T-duality, we can restrict ourselves to the topological information of the data $(P,\scG)$, that is to the equivalence classes of \v{C}ech cocycles describing the bundle $P$ and the gerbe $\scG$, without referring to their differential refinement in the form of connections. Recall that a principal circle bundle $P$ over a manifold $X$ is characterized topologically by its first Chern class $F\in \rmH^2(X,\IZ)$, whose image in de~Rham cohomology can be represented by a closed 2-form\footnote{We slightly abuse notation and denote topological classes, their images in de~Rham cohomology, and even representatives of the latter by the same symbol. Note that de~Rham cohomology misses torsion elements in \v{C}ech cohomology.} $F$ that is the curvature of a connection on $P$. Analogously, an abelian gerbe $\scG$ over a manifold $P$ is characterized topologically by its Dixmier--Douady class $H\in \rmH^3(P,\IZ)$, whose image in de~Rham cohomology can be represented by a closed 3-form $H$ which is the curvature of a connection on $\scG$, cf.~e.g.~\cite{Murray:2007ps}. The latter is often called ``$H$-flux'' in the physics literature.
  
  Mathematically, this \emph{topological T-duality}~\cite{Bouwknegt:2003vb,Bouwknegt:2003wp,Bunke:2005sn,Bunke:2005um} is due to the \emph{Gysin sequence}~\cite{Gysin:1941:61-122}, see also~\cite[Prop.~14.33]{Bott:1982aa}. Given a principal circle bundle $\check \pi:\check P\rightarrow X$ over a smooth manifold $X$ with first Chern class $\check F$, we have the following long exact sequence in cohomology:
  \begin{equation}\label{eq:Gysin_sequence}
    \ldots~\xrightarrow{~~~}~\rmH^k(X,\IZ)~\xrightarrow{~\check \pi^*~}~\rmH^k(\check P,\IZ)~\xrightarrow{~\check \pi_*~}~\rmH^{k-1}(X,\IZ)~\xrightarrow{~\check F\,\smile~}~\rmH^{k+1}(X,\IZ)~\xrightarrow{~~~}~\ldots~,
  \end{equation}
  where $\check F\,\smile$ denotes the (commutative) cup product\footnote{When considering the image in de Rham cohomology, this simply amounts to the wedge product with the 2-form curvature $\check F$.} with $\check F$. For topological T-duality, we restrict ourselves to the above window on the Gysin sequence for $k=3$. We start from a principal bundle $\check P$ over $X$ with first Chern class $\check F$ as well as an abelian gerbe $\check \scG$ over the total space of $\check P$ with Dixmier--Douady class $\check H\in \rmH^3(\check P,\IZ)$. The Gysin sequence provides us with the pushforward $\hat F=\check \pi_* H\in\rmH^2(X,\IZ)$, which we regard as the Chern class of a second principal fiber bundle $\hat \pi:\hat P\rightarrow X$ with $\check F\smile \hat F=\hat F\smile \check F=0$. Because of the latter relation, the Gysin sequence for $\hat P$ tells us that there is an $\hat H\in \rmH^3(\check P,\IZ)$, such that $\hat \pi_*\hat H=\check F$. This defines the Dixmier--Douady class of another gerbe $\hat \scG$ on the total space of $\hat P$. Altogether, topological T-duality here amounts to the exchange
  \begin{equation}
    (\check F,\check H)~~\leftrightarrow~~(\hat F,\hat H)~.
  \end{equation}
  We note that in this process, the first Chern class and the Dixmier--Douady class swap roles, and therefore the topology of the target spaces $\check P$ and $\hat P$ can be different. 
  
  The last point makes it clear that Cartesian products $X\times S^1$ are insufficient for discussing T-duality, and for general topological T-dualities, we need to provide a complete global picture of abelian gerbes on top of torus fibrations.
  
  In~\cite{Bunke:2005sn,Bunke:2005um}, topological T-dualities were described in terms of \emph{T-correspondences}, i.e.~diagram
  \begin{equation}\label{eq:T-duality_correspondence}
    \begin{tikzcd}[column sep=1cm, row sep=0.8cm]
      & & \scG_\rmC=\check\sfp^*\check\scG\otimes \hat \sfp^*\hat \scG^{-1}\cong \scI \arrow[d]& & \\
      & & \arrow[ld,"\check \sfp",swap] \check P\times_X\hat P \arrow[rd,"\hat \sfp"]& & \\
      \check\scG \arrow[r] & \check P \arrow[rd,"\check \pi"] & & \hat P \arrow[ld,"\hat \pi",swap] & \hat \scG \arrow[l]\\
      & & X & &             
    \end{tikzcd}
  \end{equation}
  where the fiber product $\check P\times_X\hat P$ is the \emph{correspondence space}, and the gerbe $\scG_\text{C}$ is isomorphic to the trivial gerbe $\scI$ with the isomorphism provided by the Poincar\'e bundle over the correspondence space. These papers also led to a precise characterization of the situation in which geometric T-dualities exist. The Serre spectral sequence associated to a torus fibration $\check \pi: \check P\rightarrow X$ defines a filtration
  \begin{equation}
    \check \pi^*\rmH^k(X)\eqqcolon F^k \subset F^{k-1} \subset \dotsb \subset F^0 \coloneqq\rmH^k(\check P)
  \end{equation}
  relating the cohomologies of the base $X$ and the total space $P$, cf.~\cite{Bunke:2005um,Baraglia:1105.0290}. This gives a classification of the Dixmier--Douady class $\check H\in \rmH^3(\check P,\IZ)$. Concretely, a background is of type $F^i$ if the 3-form image of the Dixmier--Douady class in de Rham cohomology $\check H$ vanishes after some contraction with $3-i$ vector fields along the fiber does not vanish. A geometric T-dual $(\hat P,\hat \scG)$ only exists if the gerbe $\check \scG$ has Dixmier--Douady class $\check H$ of type $F^2$ or $F^3$.
  
  \subsection{Torus bundles with \texorpdfstring{$H$}{H}-flux as principal 2-bundles}
  
  The fact that the topological input data $(\check P,\check \scG)$ for the string sigma model consists of two separate geometric structures is somewhat unsatisfying. An elegant geometric picture for topological T-duality was developed in~\cite{Nikolaus:2018qop}, where it was shown that the torus bundle $\check P$ and the abelian gerbe $\check \scG$ can be combined into a single principal 2-bundle $\check \scP$, a categorified form of a principal bundle or a ``non-abelian'' generalization of a gerbe. A T-duality can then be obtained from a span of principal 2-bundles, 
  \begin{equation}\label{eq:2-bundle-correspondence}
    \begin{tikzcd}[column sep=1cm, row sep=0.3cm]
      & \arrow[ld,"\check \sfp",swap] \scP_\rmC \arrow[rd,"\hat \sfp"]& & \\
      \check \scP & & \hat \scP
    \end{tikzcd}
  \end{equation}
  which fully subsumes the T-duality correspondences~\eqref{eq:T-duality_correspondence}. A nice feature of this construction is that the projections $\check p$ and $\hat p$ are induced from 2-group homomorphism between the structure 2-group $\underline{\sfTD}_n$ of $\scP_\rmC$ and the structure 2-group $\underline{\sfTB}_n$ of $\check \scP$ and $\hat \scP$: there is an obvious projection $\Psi:\underline{\sfTD}_n\rightarrow \underline{\sfTB}_n$, which induces the map $\check \sfp$. There is also a flip automorphism $\Phi^\text{flip}:\underline{\sfTD}_n\rightarrow \underline{\sfTD}_n$, and the composition $\Psi\circ \Phi^\text{flip}$ induces the map $\hat \sfp$.
  
  \subsection{Open questions}
  
  The geometrically appealing picture of topological T-duality evidently begs to be extended to a full version of T-duality\footnote{By ``full,'' we do not mean the inclusion of a dilaton or  Ramond--Ramond fields via twisted $K$-theory cocycles.}, involving the additional data provided by the connections on the bundle and gerbe input data $(\check P,\check \scG)$ and $(\hat P,\hat \scG)$, or, equivalently, the full Riemannian metric $G$ as well as the (locally defined) Kalb--Ramond $B$-field. It is similarly evident that we expect at least some of this data to be contained in the differential refinement, i.e.~the connections, of the principal 2-bundles $\check \scP$, $\hat \scP$, and $\scP_\text{C}$ in the correspondence~\eqref{eq:2-bundle-correspondence}. The crucial obstacle to the construction of such connections is the necessity to replace the fake-flat connections mostly used in the literature with the relatively recent concept of \emph{adjusted connections}~\cite{Kim:2019owc,Rist:2022hci}, see also \cite{Sati:2009ic,Saemann:2019dsl,Borsten:2021ljb}. As we will show, this construction reproduces indeed expected examples and features. Moreover, it was shown recently~\cite{Waldorf:2022tib} that, locally, our construction also implies the Buscher rules. The latter describe the expected relations between T-dual pairs of local metric and $B$-field data.
  
  A second question is the extension of the correspondence~\eqref{eq:2-bundle-correspondence} to non-geometric T-dualities. Already in~\cite{Nikolaus:2018qop}, half-geometric topological T-dualities were discussed; these are T-dualities involving an $F^1$-background. This was achieved by extending the structure 2-groups of the principal 2-bundles $\check \scP$ and $\scP_\rmC$ by a discrete group and removing the non-geometric leg $\hat \sfp:\scP_\rmC\rightarrow \hat \scP$. In this picture, the non-geometric background that would be described by $\hat \scP$ is resolved in a higher and doubled, but geometric background $\scP_\rmC$. 
  
  A differential refinement of this picture clearly requires more work; in particular, we expect additional scalar fields to arise, and these need to be accounted for by switching from principal 2-group bundles to principal 2-groupoid bundles.  Indeed, an elegant and physically motivated picture can be developed~\cite{Kim:2022opr}, which we will review below.
  
  \section{Categorified principal bundles with adjusted connections}
  
  Let us start with a lightning introduction to categorified principal bundles with an emphasis on adjusted connections.
  
  \subsection{Categorification}
  
  The Kalb--Ramond $B$-field is locally a 2-form on space-time and therefore describes a higher dimensional parallel transport of strings or paths along surfaces, just as local gauge potential 1-forms describe the parallel transport of a point particle along a path. Higher-dimensional parallel transport is subtle as a number of consistency conditions have to be satisfied. One such condition amounts to the following. Cut a string between two points $\bullet$ into two parts, represented by the two top arrows in the following diagram:
  \begin{subequations}\label{diag:interchange_law}
    \begin{equation}
      \begin{tikzcd}
        \bullet
        && 
        \phantom{\bullet}
        \arrow[ll,bend right=60,""{name=T1,inner sep=1pt,below}]
        \arrow[ll,""{name=M1,inner sep=1pt,below}]
        \arrow[ll,bend left=60,""{name=B1,inner sep=1pt,below}]
        \arrow[Rightarrow, from=T1,to=M1,"g_1"]
        \arrow[Rightarrow, from=M1,to=B1,"g'_1"]
        && \bullet
        \arrow[ll,bend right=60,""{name=T2,inner sep=1pt,below}]
        \arrow[ll,""{name=M2,inner sep=1pt,below}]
        \arrow[ll,bend left=60,""{name=B2,inner sep=1pt,below}]
        \arrow[Rightarrow, from=T2,to=M2,"g_2"]
        \arrow[Rightarrow, from=M2,to=B2,"g'_2"]
      \end{tikzcd}
    \end{equation}
    Then we should be able to parallel-transport this string in two steps and two different ways to the string given by the two bottom arrows: first along $g_1g_2$ and then along $g_1'g_2'$ or first along $g_2'g_2$ and then along $g_1'g_1$. The outcome should be independent of the choice of two-step parallel transport. If $g_{1,2}$ and $g'_{1,2}$ were elements in an ordinary group $\sfG$, this would impose the relation 
    \begin{equation}
      (g'_1g'_2)(g_1g_2)=(g'_1g_1)(g'_2g_2)~,
    \end{equation}
    which would force $\sfG$ to be abelian by an old argument due to Eckmann and Hilton~\cite{Hilton:1962:227-255}. If, however, we promote $g_{1,2}$ and $g'_{1,2}$ to 2-morphisms in a 2-category with horizontal and vertical compositions $\otimes$ and $\circ$, we obtain 
    \begin{equation}\label{eq:pre_interchange_law}
      (g'_1\otimes g'_2)\circ (g_1\otimes g_2)=(g'_1\circ g_1)\otimes (g'_2\circ g_2)~,
    \end{equation}
    which holds in any 2-category. In other words, an interesting compatible parallel transport along requires us to work in higher categories. 
  \end{subequations}
  
  Given a mathematical notion, we can consider its \emph{categorification}. Recall that conventional mathematical notions are defined in terms of sets, structure functions, and structure equations. Categorification then amounts to the replacement
  \begin{center}
    \begin{tabular}{rcl}
      sets & $\rightarrow$ & categories
      \\
      structure functions & $\rightarrow$ & structure functors
      \\
      structure equations & $\rightarrow$ & structure isomorphisms
    \end{tabular}
  \end{center}
  There is often a choice as to which structure isomorphisms we want to allow to be non-trivial. Furthermore, one has to carefully develop the coherence relations, i.e.~new axioms that the structure isomorphisms have to satisfy. Homomorphisms of categorified notions correspondingly consist of functors between the relevant categories, again satisfying the expected compatibility relations up to natural transformations, i.e.~morphisms between functors.
  
  Consider the example of the notion of a group. Here, we have an underlying set $\sfG$, structure maps $\circ:\sfG\times \sfG\rightarrow \sfG$ (the product), $\unit:*\rightarrow \sfG$ (the unit), and $-^{-1}:\sfG\rightarrow \sfG$ (the inverse). The structure equations capture associativity, and the relations for the unit and the inverse:
  \begin{equation}\label{eq:group_structure_equations}
    g_1\circ(g_2\circ g_3)=(g_1\circ g_2)\circ g_3~,~~~\unit\circ g_1=g_1\circ \unit=g_1~,~~~g_1\circ g_1^{-1}=g_1^{-1}\circ g_1=\unit
  \end{equation}
  for all $g_{1,2,3}\in \sfG$. In a categorified group or \emph{2-group}, we have a category $\scG=(\scG_1\rightrightarrows \scG_0)$ with $\scG_0$ the objects and $\scG_1$ the morphisms, together with a product functor $\otimes:\scG\times \scG\rightarrow \scG$, a unit $\unit:(*\rightrightarrows *)\rightarrow \scG$, and an inverse functor $\sfinv(-):\scG\rightarrow \scG$. These functors satisfy analogous relations to~\eqref{eq:group_structure_equations} up to natural transformations called associator, left- and right-unitors $\sfl$ and $\sfr$, and unit and counit:
  \begin{equation}
    g_1\otimes(g_2\otimes g_3)\Rightarrow (g_1\otimes g_2)\otimes g_3~,~~~\unit\otimes g_1\Rightarrow g_1~,~~~g_1\otimes \unit\Rightarrow g_1~,~~~\unit \Rightarrow g_1\otimes g_1^{-1}~,~~~g_1^{-1}\otimes g_1\Rightarrow \unit
  \end{equation}
  for all $g_1,g_2,g_3\in \scG_1$. 
  
  In the following, we will mostly work with \emph{strict} 2-groups, for which the above natural transformations are trivial. In these cases, we can restrict ourselves to 2-groups that have underlying categories of the form (cf.~\cite{Baez:0307200})
  \begin{equation}\label{eq:strict_Lie_2_group}
    \begin{gathered}
      \begin{tikzcd}
        \sfG\ltimes \sfH \arrow[r,shift left] 
        \arrow[r,shift right] & \sfG
      \end{tikzcd}~,~~~
      \begin{tikzcd}[column sep=2.0cm,row sep=large]
        \phantom{\sft(h)} g & \sft(h^{-1})g\arrow[l,bend left,swap,out=-20,in=200]{}{(g,h)}
      \end{tikzcd}~,
      \\
      (g_1,h_1)\circ (\sft(h_1^{-1})g_1,h_2)\coloneqq(g_1,h_1h_2)~,
    \end{gathered}    
  \end{equation}
  where $\sfG$ and $\sfH$ are groups together with an action $\acton:\sfG\times \sfH\rightarrow \sfH$ of $\sfG$ on $\sfH$. The structure functors read as 
  \begin{equation}
    \begin{aligned}
      (g_1,h_1)\otimes (g_2,h_2)&\coloneqq(g_1g_2,(g_1\acton h_2)h_1)~,
      \\
      \unit&\coloneqq(\unit_\sfG,\unit_\sfH)~,
      \\
      \sfinv(g_1,h_1)&\coloneqq (g_1^{-1},g_1^{-1}\acton h_1^{-1})~.
    \end{aligned}    
  \end{equation}    
  These strict 2-groups can alternatively be captured by \emph{crossed modules of Lie groups} $\caG=(\sfH\xrightarrow{~\sft~}\sfG,\acton)$, where $\sft:\sfH\rightarrow \sfG$ is a group homomorphism  and an action of $\sfG$ on $\sfH$ such that, for all $g\in\sfG$ and for all $h_{1,2}\in\sfH$, we have
  \begin{equation}
    \sft(g\acton h_1)\ =\ g\sft(h_1)g^{-1}
    \eand
    \sft(h_1)\acton h_2\ =\ h_1h_2h_1^{-1}~.
  \end{equation}
  
  As an example, consider the Lie 2-group $\underline{\sfTD}_n$, which will play a major role in our discussion. Its underlying groupoid is
  \begin{subequations}\label{eq:def_TD_n}
    \begin{equation}
      \begin{gathered}
        \begin{tikzcd}
          \IR^{2n}\times \IZ^{2n}\times \sfU(1)\arrow[r,shift left] 
          \arrow[r,shift right] & \IR^{2n}
        \end{tikzcd}
        \\
        \begin{tikzcd}[column sep=1.6cm,row sep=0.4cm]
          \phantom{\sft(h)} \xi & \xi-m_1\arrow[l,bend left,swap,out=-20,in=200]{}{(\xi,m_1,\phi_1)}& \xi-m_1-m_2\arrow[l,bend left,swap,out=-20,in=200]{}{(\xi-m_1,m_2,\phi_2)}\arrow[ll,bend right,out=10,in=-190]{}{(\xi,m_1+m_2,\phi_1+\phi_2)}~,
        \end{tikzcd}
        \\
        \sfid_\xi\coloneqq (\xi,0,0)~,~~~(\xi,m,\phi)^{-1}\coloneqq (\xi-m,-m,-\phi)~,
      \end{gathered}
    \end{equation}
    where $\xi\in \IR^{2n}$, $m_{1,2}\in \IZ^{2n}$ and $\phi_{1,2}\in \sfU(1)$, and the monoidal structure is given by
    \begin{equation}
      \begin{aligned}
        (\xi_1,m_1,\phi_1)\otimes (\xi_2,m_2,\phi_2)&\coloneqq(\xi_1+\xi_2,m_1+m_2,\phi_1+\phi_2-\langle \xi_1,m_2\rangle)
        \\
        \sfinv(\xi,m,\phi)&\coloneqq (-\xi,-m,-\phi-\langle \xi,m\rangle)~,
      \end{aligned}
    \end{equation}    
    where 
    \begin{equation}\label{eq:binary_map}
      \langle \xi,m\rangle \coloneqq \xi^T
      \begin{pmatrix} 
        0 & 0 \\ 
        \unit_n &0 
      \end{pmatrix}
      m~.
    \end{equation}
  \end{subequations}  
  
  \subsection{Categorified principal bundles}
  
  Essentially all definitions of principal bundles categorify, but one of the most convenient pictures providing a concrete handle on a principal bundle is given by \v{C}ech cocycles. To obtain this description, we consider a surjective submersion $\sigma:Y\rightarrow X$, and construct the corresponding \v{C}ech groupoid
  \begin{equation}
    \check\scC(Y\rightarrow X)\ \coloneqq\ \left(\begin{tikzcd}
      Y^{[2]}\arrow[r,shift left] 
      \arrow[r,shift right] & Y
    \end{tikzcd}\right)
    ~,~~~
    \begin{tikzcd}[column sep=2.0cm,row sep=large]
      y_1 \arrow[r,bend left,swap,out=-20,in=200]{}{(y_2,y_1)} & y_2\arrow[l,bend left,swap,out=-20,in=200]{}{(y_1,y_2)}~,
    \end{tikzcd}
  \end{equation}
  where $Y^{[k]}\coloneqq\{(y_1,\ldots,y_k)|\sigma(y_1)=\ldots=\sigma(y_k)\}$ is the fiber product over $X$. In most cases, $Y$ is the atlas of a manifold, and $Y^{[2]}$ contains the non-empty intersections. The \v{C}ech cocycle for a principal $\sfG$-bundle subordinate to the surjective submersion $\sigma$ is then simply a functor 
  \begin{equation}
    \Phi:\check\scC(Y\rightarrow X)\rightarrow \sfB\sfG~,
  \end{equation}
  where $\sfB\sfG=(\sfG\rightrightarrows*)$ is the one-object groupoid with the group $\sfG$ as its set of morphisms. Clearly, $\Phi$ is trivial on objects, and on morphisms it amounts to a map $g:Y^{[2]}\rightarrow \sfG$ which satisfies
  \begin{equation}
    g(y_1,y_2)g(y_2,y_3)=g(y_1,y_3)~,
  \end{equation}
  the cocycle relation for transition functions $g$ of a principal $\sfG$-bundle. One can also show that bundle isomorphisms (i.e.~physicists' gauge transformations) arise from the corresponding natural transformations. 
  
  This picture readily generalizes: we can trivially regard the \v{C}ech groupoid $\check\scC(Y\rightarrow X)$ as a 2-groupoid, and every 2-group $\scG$ comes with a one-object Lie 2-groupoid $\sfB\scG$. A principal 2-bundle is then a 2-functor\footnote{Even if the gauge 2-group is strict, we allow for weak 2-functors (sometimes called pseudofunctors).}
  \begin{equation}
    \Phi:\check\scC(Y\rightarrow X)\rightarrow \sfB\scG~,
  \end{equation}
  and gauge transformations are natural 2-transformations between these. 
  
  In the case of a strict Lie 2-group of the form~\eqref{eq:strict_Lie_2_group}, the corresponding 2-functor amounts to data 
  \begin{subequations}\label{eq:geom_cocyles}
    \begin{equation}
      h\in C^\infty(Y^{[3]},\sfH)\eand g\in C^\infty(Y^{[2]},\sfG)
    \end{equation}
    satisfying
    \begin{equation}
      \begin{aligned}
        h_{ikl}h_{ijk}\ &=\ h_{ijl}(g_{ij}\acton h_{jkl})~,
        \\
        g_{ik}\ &=\ \sft(h_{ijk})g_{ij}g_{jk}
      \end{aligned}
    \end{equation}
    for all $(ijk)\in Y^{[3]}$ and $(ij)\in Y^{[2]}$, where we have abbreviated $g(y_i,y_j)=g_{ij}$, etc.
  \end{subequations}  
  
  We note that, up to technical difficulties, higher generalizations to principal $n$-bundles are straightforward. Moreover, it is also immediately clear how to define principal groupoid and 2-groupoid bundles in terms of functors.
  
  \subsection{Fake-flat connections}
  
  The definition of higher connections is somewhat more involved. One can define them from a number of different perspectives; see e.g.~the approaches in the original literature~\cite{Breen:math0106083}, \cite{Aschieri:2003mw}. The differential refinement of the cocycle data~\eqref{eq:geom_cocyles} found in these papers consists of the following set of maps for a 2-group of the form~\eqref{eq:strict_Lie_2_group}:
  \begin{subequations}\label{eq:unadjusted_diff_refinement}
    \begin{equation}
      \begin{aligned}
        h&\in \Omega^0(Y^{[3]},\sfH)~, & \Lambda&\in \Omega^1(Y^{[2]},\frh)~, & B&\in \Omega^2(Y,\frh)~, &\delta\in \Omega^2(Y^{[2]},\frh)~,
        \\
        g&\in \Omega^0(Y^{[2]},\sfG)~, & A&\in \Omega^1(Y,\frg)~, & & 
      \end{aligned}
    \end{equation}
    where $\frg$ and $\frh$ are the Lie algebras of the Lie groups $\sfG$ and $\sfH$, respectively. We clearly see the usual pattern, familiar e.g.~from Deligne cohomology or simply the \v{C}ech--de Rham correspondence, that one can trade one \v{C}ech-degree for a de Rham-degree. Only the datum $\delta$ does not fit this pattern. Studies of the resulting notion of higher parallel transport~\cite{Baez:0511710,Schreiber:0705.0452,Schreiber:2008aa} suggested that $\delta$ is spurious, and it has been dropped in much of the work on higher gauge theory, following~\cite{Baez:0511710}. In addition to~\eqref{eq:geom_cocyles}, one then has the cocycle relations
    \begin{equation}\label{eq:unadjusted_diff_refinement_conditions}
      \begin{aligned}
        \Lambda_{ik}\ &=\ \Lambda_{jk}+g_{jk}^{-1}\acton\Lambda_{ij}-g_{ik}^{-1}\acton(h_{ijk}\nabla_i h_{ijk}^{-1})~,
        \\
        A_j\ &=\ g^{-1}_{ij}A_ig_{ij}+g^{-1}_{ij}\rmd g_{ij}-\sft(\Lambda_{ij})~,
        \\
        B_j\ &=\ g^{-1}_{ij}\acton B_i+\rmd\Lambda_{ij}+A_j\acton \Lambda_{ij}+\tfrac12[\Lambda_{ij},\Lambda_{ij}]
      \end{aligned}
    \end{equation}
    for all $(ijk)\in Y^{[3]}$ and $(ij)\in Y^{[2]}$.
  \end{subequations} 
  Contrary to ordinary principal bundles, the consistency condition that the local $B$-field components glue correctly together over $Y^{[3]}$ leads to the constraint 
  \begin{equation}
    (g_{jk}^{-1}g_{ij}^{-1})\acton (h_{ijk}^{-1}(\caF_i\acton h_{ijk})\stackrel{!}{=}0
  \end{equation}
  over $Y^{[3]}$, where 
  \begin{equation}
    \caF_i\coloneqq \rmd A_i+\tfrac12[A_i,A_i]+\sft(B_i)
  \end{equation}
  is the so-called \emph{fake curvature}. This is guaranteed if we impose \emph{fake flatness}, i.e.~the condition $\caF_i=0$. As shown in~\cite{Schreiber:0705.0452,Schreiber:2008aa}, fake flatness is not only sufficient but also necessary for rendering the corresponding higher parallel transport reparameterization-invariant. Finally, the gauge transformation of the non-abelian 3-form curvature
  \begin{equation}\label{eq:3_form_curvature_unadjusted}
    H_i\coloneqq \rmd B_i+A_i\acton B_i
  \end{equation}
  is of the form
  \begin{equation}
    H_i\rightarrow \tilde H_i=g_i\acton H_i-\caF_i\acton \Lambda_i
  \end{equation}
  for gauge parameters $g\in \Omega^0(Y,\sfG)$ and $\Lambda\in \Omega^1(Y,\frh)$. This implies that the self-duality equation $H=\star H$ believed to be a vital ingredient in 6d superconformal field theories (cf.~e.g.~\cite{Rist:2020uaa} and references therein) is only covariant for fake-flat connections.
  
  Unfortunately, the condition $\caF_i=0$ implies that, locally, the gauge potential 1-form $A_i$ can be gauged away~\cite{Saemann:2019dsl} (see also~\cite{Gastel:2018joi}), which in turn reduces the principal 2-bundle to an abelian gerbe. This situation is simply too restrictive for most applications in physics.
  
  \subsection{Adjusted connections}
  
  In a special case and at an infinitesimal level, a solution to this problem, but formulated in a much less mathematical language, had been known for a long time~\cite{Bergshoeff:1981um,Chapline:1982ww}, and this was developed mathematically in the context of higher gauge theory in~\cite{Sati:2008eg,Waldorf:2009uf,Sati:2009ic} for the case of the skeletal string Lie 2-algebra. This picture was then extended in~\cite{Saemann:2017rjm,Saemann:2019dsl,Saemann:2019dsl,Kim:2019owc,Borsten:2021ljb}, and the complete, global definition of the notion of adjusted connection including finite gauge transformations was given in~\cite{Rist:2022hci}. In this picture, the definition of the 3-form curvature is modified, which implies a change in gauge transformations and cocycle relations that allows to drop the fake flatness condition consistently.
  
  As a perhaps familiar example from physics, let us consider a slight generalization of the infinitesimal, local adjustment encountered in heterotic supergravity\footnote{The explicit formulas for this example were derived already in~\cite{Bergshoeff:1981um,Chapline:1982ww}, albeit without reference to gerbes or higher algebra.}. For each metric (i.e.~quadratic) Lie algebra $(\frg,(-,-))$, there is a corresponding skeletal string Lie 2-algebra, a 2-term $L_\infty$-algebra whose explicit form is irrelevant here. It is important that, on a patch $U$, the resulting gauge potentials are of the form
  \begin{equation}
    A\in~\Omega^1(U,\frspin(n))\eand B\in\Omega^2(U)~,
  \end{equation}
  and the curvatures read as 
  \begin{equation}
    \begin{aligned}
      F&\coloneqq \rmd A+\tfrac12 [A,A]
      \\
      H&\coloneqq \rmd B-\tfrac{1}{3!}(A,[A,A])+(A,F)
      \\
      &\phantom{:}= \rmd B+(A,\rmd A)+\tfrac13 (A,[A,A])=\rmd B+\text{cs}(A)~.
    \end{aligned}
  \end{equation}
  Here, the term $(A,F)$ is the adjustment of the unadjusted curvature $H=\rmd B-\tfrac1{3!}(A,[A,A])$, and we obtain the familiar Bianchi identities:
  \begin{equation}
    \rmd F+[A,F]=0~,~~~\rmd H=(F,F)~.
  \end{equation}
  
  The above adjustment is hard to integrate explicitly, and it is beneficial to switch to the equivalent strict model, as done in~\cite{Saemann:2017rjm,Saemann:2019dsl}. The resulting adjustment induces an adjustment of the cocycle condition for $B$ in~\eqref{eq:unadjusted_diff_refinement_conditions}, and for a generic strict gauge 2-group, this modification takes the form~\cite{Rist:2022hci}
  \begin{equation}\label{eq:adjusted_cocycle_relation}
    B_j=g^{-1}_{ij}\acton B_i+\rmd\Lambda_{ij}+A_j\acton \Lambda_{ij}+\tfrac12[\Lambda_{ij},\Lambda_{ij}]-\kappa(g_{ij},F_i)~.
  \end{equation}
  Here, $\kappa:\sfG\times \frg\rightarrow \frh$ is the \emph{adjustment datum}, a map linear in $\frg$, which satisfies the relation
  \begin{equation}\label{eq:adjustment_condition}
    \begin{aligned}
      (g_2^{-1}g_1^{-1})\acton (h^{-1}(X\acton h))
      &+g_2^{-1}\acton\kappa(g_1,X)
      \\
      &+\kappa(g_2,g_1^{-1}X g_1-\sft(\kappa(g_1,X)))-\kappa(\sft(h)g_1g_2,X)\ =\ 0
    \end{aligned}
  \end{equation}
  for all $g_1,g_2\in \sfG$, $h\in \sfH$, and $X\in \frg$~\cite{Rist:2022hci}. 
  
  The most important example of an adjustment, namely an adjustment for the strict 2-group model of the string group, was given in given~\cite{Rist:2022hci}. Another large class of (infinitesimal) adjustments, also for principal $n$-bundles with $n>2$, was found in the context of the tensor hierarchies of gauged supergravity~\cite{Borsten:2021ljb}. In this case, the algebraic origin of the adjustment datum is particularly clear.
  
  For our discussion, we are mostly in interested in the strict 2-group $\underline{\sfTD}_n$ defined in~\eqref{eq:def_TD_n}. A corresponding adjustment is given by the pairing~\eqref{eq:binary_map}, extended to 
  \begin{equation}\label{eq:adjustment_TD_n}
    \begin{aligned}
      \kappa: \IR^{2n}\times \IR^{2n}&\rightarrow \sfU(1)~,
      \\
      (\xi_1,\xi_2)&\mapsto \xi^T_1
      \begin{pmatrix} 
        0 & 0 \\ 
        \unit_n &0 
      \end{pmatrix}
      \xi_2~,
    \end{aligned}
  \end{equation}
  and one readily verifies that $\kappa$ satisfies~\eqref{eq:adjustment_condition}. Just as in the case of the 2-group model of the string group, it is remarkable that the adjustment datum is already part of the description of the 2-group $\underline{\sfTD}_n$ itself.
  
  \subsection{Non-trivial example: higher non-abelian instantons}
  
  To reassure the reader that the above theory indeed comes with non-trivial and interesting examples, let us briefly sketch the main example of~\cite{Rist:2022hci}, see also~\cite{Roberts:2022wwl}, where the higher analogue of an instanton was constructed. Recall that the fundamental $\sfSU(2)$-instanton on $S^4$ is described by the Hopf fibration
  \begin{equation}
    S^3\hookrightarrow S^7\rightarrow S^4~.
  \end{equation}
  We can now double the gauge group to $\sfSpin(4)\cong \sfSU(2)\times \sfSU(2)$ to obtain the principal $\sfSpin(4)$-bundle
  \begin{equation}
    \sfSpin(5)\rightarrow \sfSpin(5)/\sfSpin(4)\cong S^4~.
  \end{equation}
  The Maurer--Cartan form on $\sfSpin(5)$ induces a connection on this bundle, which splits into the two $\frsu(2)$ components. This connection is a pair of a fundamental instanton and a fundamental anti-instanton. The contribution of these two components to the first Pontryagin class are equal but of opposite sign, an important condition for being able to lift this $\sfSpin(4)$-bundle to a $\sfString(4)$-bundle. The latter is a 2-group which can be regarded as a particular categorified central extension of $\sfSpin(4)$.
  
  The resulting $\sfString(4)$-bundle can be seen as a 2-group coset space, which fibers over $S^4$, trivially regarded as a categorified or 2-space:
  \begin{equation}
    \sfString(5)\rightarrow \sfString(5)/\sfString(4)\cong (S^4\rightrightarrows S^4)~,
  \end{equation}
  and the connection on this space requires adjustment~\cite{Rist:2022hci}. This principal 2-bundle can be seen as the higher analogue of an instanton as it is literally the lift of an instanton bundle to a 2-group bundle. On the other hand, it also describes a non-abelian self-dual string in the sense of~\cite{Saemann:2017rjm}, the higher analogue of a monopole.
  
  \section{Geometric T-duality with principal 2-bundles}
  
  After this lengthy prelude, let us come to the description of T-duality with principal 2-bundles, starting from the picture of~\cite{Nikolaus:2018qop}.
  
  \subsection{Topological T-duality from principal 2-bundles}
  
  Categorified principal bundles over a manifold $X$ with structure 2-group $\underline{\sfTB_n}$ are in one-to-one correspondence with geometric T-backgrounds, i.e.~abelian gerbes on principal torus fibrations over $X$~\cite{Nikolaus:2018qop}. The crossed module of the 2-group $\underline{\sfTB}_n$ is given by 
  \begin{equation}\label{eq:def_TB_F2}
    \begin{gathered}
      \sfTB_n~=~\big(\IZ^n\times C^\infty(\IT^n,S^1)\xrightarrow{~\sft~}\IR^n\big)~,
      \\
      \sft(m,f)=m~,~~~\xi\acton(m,f)=(m,c\mapsto f(c-\xi))=(m,f\circ \sfs_{\xi})
    \end{gathered}
  \end{equation}
  for all $\xi\in \IR^n$, $m\in \IZ^n$, and $f\in C^\infty(\IT^n,S^1)$, where $\sfs_\xi$ denotes the function $\sfs_\xi t\mapsto t-\xi$. 
  
  T-duality along the torus directions is then a span of principal 2-bundles 
  \begin{equation}\label{eq:geometric_T_duality_span}
    \begin{tikzcd}
      & \arrow[ld,"\check \sfp",swap] \scP_\rmC \arrow[rd,"\hat \sfp"]& & \\
      \check \scP & & \hat \scP
    \end{tikzcd}
  \end{equation}
  where $\check \scP$ and $\hat \scP$ are principal $\underline{\sfTB}_n$-bundles, encoding geometric T-backgrounds. The 2-bundle $\scP_\rmC$, however, is a principal $\underline{\sfTD}_n$-bundle. The cocycle descriptions of these principal 2-bundles are obtained from specializations of the cocycles~\eqref{eq:geom_cocyles}. For principal $\underline{\sfTD}_n$-bundles, we have 
  \begin{subequations}\label{eq:TD-top-cocycles}
    \begin{equation}
      h=(m_{ijk},\phi_{ijk})\in C^\infty(Y^{[3]},\IZ^{2n}\times\sfU(1))\eand 
      g=(\xi_{ij})\in C^\infty(Y^{[2]},\IR^{2n})
    \end{equation}
    with $(ij)\in Y^{[2]}$ and $(ijk)\in Y^{[3]}$, which satisfy\footnote{We use additive notation for the group $\sfU(1)$.}
    \begin{equation}
      \begin{aligned}
        \phi_{ikl}+\phi_{ijk}&=\phi_{ijl}+\phi_{jkl}-\langle \xi_{ij},m_{jkl}\rangle~,
        \\
        m_{ikl}+m_{ijk}&=m_{ijl}+m_{jkl}~,
        \\
        \xi_{ik} &= m_{ijk}+\xi_{ij}+\xi_{jk}
      \end{aligned}
    \end{equation}
    on $Y^{[4]}$ and $Y^{[3]}$, respectively. We thus see that a principal $\underline{\sfTD}_n$-bundle $\scP$ contains a principal $\sfU(1)^{2n}$-bundle $\scP^\circ$. Because $\sfU(1)^{2n}$ is identified with $\IR^{2n}/\IZ^{2n}$, the cocycle $(m,\xi)$ takes the form of a principal 2-bundle for a structure 2-group given by a crossed module of the form $(\IZ^{2n}\hookrightarrow\IR^{2n})$.
  \end{subequations}    
  
  Note that our description of principal $2$-bundles as functors makes it evident that a morphism $\Phi:\scG_1\rightarrow \scG_2$ between 2-groups $\scG_{1,2}$ induces a bundle morphism from a principal $\scG_1$-bundle to a principal $\scG_2$-bundle. For T-duality, we have the strict 2-group homomorphism\footnote{See the appendix of~\cite{Kim:2022opr} for details.}
  \begin{equation}
    \Psi : \underline{\sfTD}_n \rightarrow \underline{\sfTB}_n~,~~~
    \Psi_1\left(
    \begin{pmatrix} 
      \hat \xi\\ \check \xi
    \end{pmatrix}
    ,
    \begin{pmatrix} 
      \hat m\\ \check m
    \end{pmatrix}
    ,\phi
    \right)=(\check \xi,\check m,c\mapsto \phi+\hat m^\rmT  \check gc)~,
  \end{equation}
  which induces the projection $\check \sfp$. On the other hand, there is a (weak) 2-group automorphism
  \begin{subequations}\label{eq:flip_automorphism}
    \begin{equation}
      \Phi^\text{flip}:\underline{\sfTD}_n\rightarrow \underline{\sfTD}_n~,
    \end{equation}
    which in particular interchanges the two $n$-dimensional components in $\IR^{2n}$ and $\IZ^{2n}$:
    \begin{equation}
      \Phi^\text{flip}\left(
      \begin{pmatrix} 
        \hat \xi\\ \check \xi
      \end{pmatrix}
      ,
      \begin{pmatrix} 
        \hat m\\ \check m
      \end{pmatrix}
      ,\phi
      \right)=\left(
      \begin{pmatrix} 
        \check \xi\\ \hat \xi
      \end{pmatrix}
      ,
      \begin{pmatrix} 
        \check m\\ \hat m
      \end{pmatrix}
      ,\ldots
      \right)~,
    \end{equation}
  \end{subequations}  
  and the projection $\hat \sfp$ is induced by the concatenation $\Psi\circ \Phi^\text{flip}$.
  
  \subsection{Differential refinement of principal \texorpdfstring{$\underline{\sfTD}_n$}{TDn}-bundles}
  
  Let us directly come to the differential refinement of this picture. The explicit form of differentially refined cocycles for principal $\underline{\sfTD}_n$-bundles is given by~\eqref{eq:geom_cocyles} and~\eqref{eq:unadjusted_diff_refinement}, where the unadjusted cocycle relation for the $B_i$ is replaced by its adjusted form~\eqref{eq:adjusted_cocycle_relation} with the adjustment given in~\eqref{eq:adjustment_TD_n}. The flip automorphism~\eqref{eq:flip_automorphism} of the 2-group $\underline{\sfTD}_n$ induces an action on the differentially refined cocycle data.
  
  Interestingly, the strict structure 2-group $\underline{\sfTB}_n$ of the principal 2-bundles $\check \scP$ and $\hat \scP$ in the span of 2-bundles~\eqref{eq:geometric_T_duality_span} does \emph{not} admit an adjustment, as one sees after specializing~\eqref{eq:adjustment_condition} to this case~\cite{Kim:2022opr}. However, it was shown in~\cite[Remark 3.4.6]{Nikolaus:2018qop} that the classifying spaces $\sfB\underline{\sfTD}_n$ and $\sfB\underline{\sfTB}_n$ are equivalent, which means that for any principal $\underline{\sfTB}_n$-bundle, there is an equivalence class of principal $\underline{\sfTD}_n$-bundles. In this sense, it is sufficient to restrict ourselves to $\underline{\sfTD}_n$-bundles and to study T-duality purely in terms of those. For non-geometric T-dualities, this is necessary in any case, even without differential refinement. 
  
  Losing the two legs $\check \sfp$ and $\hat \sfp$ in the correspondence~\eqref{eq:geometric_T_duality_span} is not a problem as one can still recover the data of two principal torus bundles with gerbes on their total spaces from a principal $\underline{\sfTD}_n$-bundle. This is clear for the topological cocycles as we could simply restrict to topological data an use~\eqref{eq:geometric_T_duality_span}. There is, however, a more elegant method. We can pull back $\scP$ along the projection $\scP^\circ \rightarrow X$ of the principal $\sfU(1)^{2n}$-bundle $\scP^\circ$ that $\scP$ contains, which then gives us explicit cocycles for a torus fibration with connection and an abelian gerbe on its total space with connection~\cite{Kim:2022opr}. A second background is obtained in this fashion after first applying the flip automorphism~\eqref{eq:flip_automorphism} to the differentially refined cocycle data. 
  
  \subsection{The T-duality 2-group \texorpdfstring{$\scGO(n,n;\IZ)$}{GO(n,n;Z)} and its manifest action}
  
  While T-duality is often explained in terms of a $\IZ_2$-action or involution, this picture is incomplete even in the case of a single T-duality direction. Generically, for T-duality along an $n$-dimensional torus $T^n$, there is the discrete group $\sfO(n,n;\IZ)$ of transformations acting on the T-background data leading to physically equivalent sigma models, cf.~\cite{Giveon:1988tt,Shapere:1988zv}. 
  
  Recall that a successful strategy for studying symmetries is to make them manifest. Making the T-duality group $\sfO(n,n;\IZ)$ manifest for T-backgrounds is one of the goals of \emph{double field theory}; see~\cite{Aldazabal:2013sca,Berman:2013eva,Hohm:2013bwa} for reviews. As we will explain below, the principal $\underline{\sfTD}_n$-bundles perform the same task for principal $\underline{\sfTB}_n$-bundles and hence for general T-backgrounds.
  
  First, we note that the group $\sfO(n,n;\IZ)$ is still not general enough since orientation reversal requires extension to the group $\sfGO(n,n;\IZ)=\sfO(n,n;\IZ)\rtimes \IZ_2$, where the nontrivial element of $\IZ_2$ acts as conjugation by $\left(\begin{smallmatrix}0&\unit_n\\-\unit_n&0\end{smallmatrix}\right)$. 
  
  There is now an evident notion of the automorphism 2-group $\sfAut(\scG)$ of a 2-group $\scG$, which allows us to define the action of a 2-group $\scH$ on $\scG$ as a 2-group homomorphism $\Phi:\scH\rightarrow \sfAut(\scG)$,~cf.~\cite{Garzn:2001aa}. The automorphism 2-group of $\underline{\sfTD}_n$ was characterized as a non-central extension of $\sfGO(n,n;\IZ)$ in~\cite{Waldorf:2022lzs}, and an explicit description as a weak 2-group was given in~\cite{Kim:2022opr}. As observed in~\cite{Kim:2022opr}, neither the group $\sfGO(n,n;\IZ)$ nor the group $\sfO(n,n;\IZ)$, trivially regarded as 2-groups, admit an action on $\underline{\sfTD}_n$ with the required properties. This implies that the T-duality group $\sfGO(n,n;\IZ)$ has to be replaced by the T-duality 2-group $\scGO(n,n;\IZ)$.
  
  This 2-group now acts directly on $\underline{\sfTD}_n$, making T-duality manifest. The action further induces an action on principal $\underline{\sfTD}_n$-bundles by postcomposing the functors defining the latter with the 2-group action. Preserving this $\scGO(n,n;\IZ)$-covariance is a useful constraint when further extending the above description by scalar fields. Moreover, $\scGO(n,n;\IZ)$ contains the flip automorphism~\eqref{eq:flip_automorphism}; thus having an $\scGO(n,n;\IZ)$-action indicates the action of T-dualities.
  
  \subsection{Scalar fields}
  
  Recall the relation between the Gysin sequence~\eqref{eq:Gysin_sequence} and topological T-duality. The first Chern class of the circle bundle in the T-dual T-background was obtained by a pushforward of the Dixmier--Douady class of the abelian gerbe in the original T-background. At the level of differential forms, this amounts to an integration of the gerbe curvature $\check H$ along the fiber direction, leading to the curvature $\hat F$ of the T-dual circle bundle. More physically, this is simply a Kaluza--Klein (KK) reduction in the T-duality directions. 
  
  Clearly, KK reducing a 2-form $B$ along the circle fiber yields a 1-form $A$. Similarly, a KK reduction of the metric yields a second 1-form $A$ together with a scalar field $\phi$. We also note that KK reductions along several circle directions will produce additional scalar fields also from the 2-form $B$. For a full geometric T-duality between input data of sigma models, it therefore remains to consistently incorporate all these scalar fields. 
  
  Fortunately, there is a clear physical description of the target space of the scalar fields, namely the \emph{Narain moduli space}~\cite{Narain:1985jj}, the moduli of the Riemannian metric and the Kalb--Ramond $B$-field on $T^n$. This space is given by
  \begin{equation}
    M_n=\sfO(n,n;\IZ)~\backslash~\sfO(n,n;\IR)~/~\big(\sfO(n;\IR)\times \sfO(n;\IR)\big)=\sfO(n,n;\IZ)~\backslash~Q_n
  \end{equation}
  with $Q_n\coloneqq \sfO(n,n;\IR)~/~\big(\sfO(n;\IR)\times \sfO(n;\IR)\big)$. As explained above, we should really replace $\sfO(n,n;\IZ)$ with $\sfGO(n,n;\IZ)$ to allow for orientation reversing. Also, we follow the strategy familiar from gauge theory of resolving the quotient $\sfGO(n,n;\IZ)\backslash Q_n$ in its corresponding action groupoid. That is, we consider the groupoid
  \begin{equation}\label{eq:scalar_groupoid}
    \left(\begin{tikzcd}
      \sfGO(n,n;\IZ)\ltimes Q_n\arrow[r,shift left] 
      \arrow[r,shift right] & Q_n
    \end{tikzcd}\right)
    ~,~~~
    \begin{tikzcd}[column sep=2.0cm,row sep=large]
      q\arrow[r,bend left,swap,out=-30,in=200]{}{(g,q)^{-1}} & g^{-1}\acton q\arrow[l,bend right,swap,out=-20,in=210]{}{(g,q)}~,
    \end{tikzcd}
  \end{equation}
  and the quotient space $\sfGO(n,n;\IZ)\backslash Q_n$ is given by the isomorphism classes in this groupoid. 
  
  The scalar groupoid~\eqref{eq:scalar_groupoid} can now be nicely combined with the 2-group $\underline{\sfTD}_n$ into a 2-groupoid: first, we trivially extend the group $\sfGO(n,n;\IZ)$ to the T-duality 2-group $\scGO(n,n;\IZ)$. Then we extend the morphisms for every $q\in Q_n$ by a copy of $\underline{\sfTD}_n$ and have $\sfGO(n,n;\IZ)$ act diagonally on both $Q_n$ and $\underline{\sfTD}_n$. The resulting Lie 2-groupoid $\scTD_n$ has the following 2-, 1-, and 0-cells:
  \begin{equation}
    \begin{aligned}
      (\scTD_n)_2&=\sfGO(n,n;\IZ)\times \IZ^{2n}\times\IR^{2n}\times \IZ^{2n}\times \sfU(1)\times Q_n~,
      \\
      (\scTD_n)_1&=\sfGO(n,n;\IZ)\times \IR^{2n}\times Q_n~,
      \\
      (\scTD_n)_0&=Q_n~,
    \end{aligned}~~~
    \begin{tikzcd}[column sep=3.8cm,row sep=large]
      q & \ar[l,bend left=45, "{(g,\xi-m,q)}", ""{name=U,inner sep=1pt,above}] \ar[l, xshift=0.09cm,bend right=45, "{~~(g,\xi,q)}", swap, ""{name=D,inner sep=1pt,below}] 
      g^{-1}\acton q \arrow[Rightarrow,from=U, to=D, "{(g,\xi,z,m,\phi,q)}",swap,sloped]
    \end{tikzcd}
  \end{equation}
  In the 2-cells, the factor $\sfGO(n,n;\IZ)\times \IZ^{2n}$ arises from $\scGO(n,n;\IZ)$, and the factor $\IR^{2n}\times \IZ^{2n}\times \sfU(1)$ stems from the copy of $\underline{\sfTD}_n$.
  
  Our definition of categorified principal bundles now readily extends to principal $\scTD_n$-bundles: these are weak 2-functors from the \v{C}ech groupoid $\check\scC(Y\rightarrow X)$ of some surjective submersion $Y\rightarrow X$ to $\scTD_n$. Similarly, it is not hard to extend the definitions of adjusted connection, see~\cite{Kim:2022opr}. 
  
  \subsection{Full geometric T-duality}\label{ssec:full_T_duality}
  
  From our perspective, a full\footnote{again neglecting the dilaton and Ramond--Ramond-fields} geometric T-duality is now  a \emph{geometric} principal $\scTD_n$-bundle $\scP$ over $X$ with adjusted connection. By geometric, we mean here such principal $\scTD_n$-bundles, for which the $\sfGO(n,n;\IZ)$-valued component of the cocycle takes values in the geometric subgroup $\sfGL(n;\IZ)$ of the T-duality group $\sfGO(n,n;\IZ)$. 
  
  We note that we have an explicit action of the group $\sfGO(n,n;\IZ)$ on these bundles, that among other transformations also induces the action of the T-duality involution by the flip automorphism~\eqref{eq:flip_automorphism}. To make this action manifest, it is convenient to arrange the scalar fields originating from KK reduction of the metric $G$ and the $B$-field of a T-background in the form of the \emph{generalized metric}
  \begin{equation}\label{eq:generalized_metric}
    \caH\coloneqq \begin{pmatrix}
      G-BG^{-1} B & B G^{-1}
      \\
      -G^{-1} B & G^{-1}
    \end{pmatrix}~,
  \end{equation}
  as usually done in generalized geometry and double field theory. The action of $\sfGO(n,n;\IZ)$ on the scalar fields is then simply the adjoint action on $\caH$.
  
  The crucial test for this description of T-duality is certainly that it reproduces the Buscher rules. The corresponding cumbersome computation was performed in~\cite{Waldorf:2022tib}, where it was shown that this expectation is met.
  
  Instead of giving the picture we sketched above in all details (which can be found in~\cite{Kim:2022opr}), let us consider the instructive example of T-dualities with T-backgrounds given by abelian gerbes $\scG_\ell$ over three-dimensional nilmanifolds $N_k$. The latter are principal circle bundles over the 2-torus $T^2$ characterized by their first Chern numbers $k\in \rmH^2(T^2,\IZ)\cong \IZ$. Subordinate to the convenient surjective submersion $Y=\IR^2\rightarrow T^2$, we have the differentially refined cocycle
  \begin{equation}
    \begin{gathered}
      g\in \Omega^0(Y^{[2]},\sfU(1))\eand A\in \Omega^1(Y,\fru(1))~,
      \\
      g(x,y;x',y')\coloneqq k(x'-x)y\eand 
      A(x,y)\coloneqq k x\,\rmd y~,
    \end{gathered}
  \end{equation}
  In addition, we have a scalar field $\psi:Y\rightarrow \IR^+$ that parameterizes the size of the circle direction. For simplicity, we assume a constant scalar field,
  \begin{equation}
    \psi=2\pi R~.
  \end{equation}
  The above data combines into the usual Kaluza--Klein metric $G_{\text{KK}}$ for $N_k$ on $T^2$,
  \begin{equation}
    G_{\text{KK}}=\pi^*G_{T^2}+\psi^2 A\odot A~,
  \end{equation}
  where $\pi:N_k\rightarrow T^2$ is the bundle projection and $G_{T^2}$ is the usual flat metric on $T^2$ with radii $1$.
  
  The Dixmier--Douady class of $\scG_\ell$ is an element $\ell\in \rmH^3(N_k,\IZ)\cong \IZ$. Subordinate to the surjective submersion\footnote{The nilmanifold $N_k$ can also be seen as the quotient of $\IR^3$ by the relations
    \begin{equation}
      (x,y,z)\sim(x,y+1,z)\sim(x,y,z+1)\sim(x+1,y,z-ky)~.
    \end{equation}} $Z=\IR^3\rightarrow N_k$, the abelian gerbe is described by the differentially refined cocycle
  \begin{equation}
    \begin{gathered}
      h\in \Omega^0(Z^{[3]},\sfU(1))~,~~~\Lambda \in \Omega^1(Z^{[2]},\fru(1))~,\eand B\in \Omega^2(Z,\fru(1))~,
      \\
      h(x,y,z;x',y',z';x'',y'',z'')\coloneqq\ell (x-x')(y'-y'')z~,
      \\
      \Lambda(x,y,z;x',y',z') =\ell(x-x')y \rmd z~,\eand
      B(x,y,z)= \ell x\,\rmd y\wedge\rmd z~.
    \end{gathered}
  \end{equation}
  A T-duality along the fiber direction in $N_k$ leads to another gerbe on a nilmanifold with an interchange of the topological invariants and an inversion of the fiber radius:
  \begin{equation}
    (N_k,\scG_\ell,\psi)\leftrightarrow (N_\ell,\scG_k,\psi^{-1})~.
  \end{equation}
  
  To describe this T-duality in terms of higher geometry, we work with principal $\scTD_1$-bundles (there is one T-duality direction) over $T^2$. Subordinate again to $Y=\IR^2\rightarrow T^2$, we have the differentially refined cocycles~\cite{Kim:2022opr}
  \begin{subequations}\label{eq:nilmanifold_cocycle}
    \begin{equation}
      \begin{gathered}
        \phi(x,y;x',y';x'',y'')= \tfrac12k\ell\left(y'(xx''-xx'-x'x'')-(x''-x')(y'^2-y^2)x\right)~,
        \\
        m(x,y;x',y';x'',y'')= \begin{pmatrix}
          -\ell(x''-x')(y'-y)
          \\
          -k(x''-x')(y'-y)
        \end{pmatrix}~,
        \eand 
        g(x,y;x',y')=\begin{pmatrix}
          \ell(x'-x)y
          \\
          k(x'-x)y
        \end{pmatrix}~,
        \\
        \Lambda(x,y;x',y') = \tfrac12k\ell(xx'\,\rmd y+(xy+x'y'+y^2(x'-x))\,\rmd x)~,
        \\
        A=\begin{pmatrix}
          k x\,\mathrm dy
          \\
          \ell x\,\mathrm dy
        \end{pmatrix}~,\eand B(x,y) = 0~,
        \\
        \psi(x)=\begin{pmatrix}
          2\pi R & 0 
          \\
          0 & \frac{1}{2\pi R}
        \end{pmatrix}~,
      \end{gathered}
    \end{equation}
    where we identified the scalar field in $Q_1\coloneqq\sfO(1,1;\IR)/(\sfO(1,\IR)\times \sfO(1,\IR))\cong \IR^+$ with the corresponding generalized metric~\eqref{eq:generalized_metric}.
  \end{subequations}    
  We note that all expressions are explicit and relatively simple. Furthermore, the topological charges $k$ and $\ell$ appear completely on equal footing, as do the fiber radius and its inverse. A T-duality is then described by an action of the flip automorphism~\eqref{eq:flip_automorphism} on this data. Following our above prescription of recovering the individual T-backgrounds then yields indeed the pair $(N_k,\scG_\ell,\psi)$ and $(N_\ell,\scG_k,\frac{1}{\psi})$.
  
  \section{Non-geometric T-dualities}
  
  There are two steps in the generalization of geometric T-dualities with $F^2$-backgrounds to general backgrounds. The first one involves \emph{T-folds}, which are spaces that are locally abelian gerbes over ordinary manifolds but whose local data is glued together by T-duality transformations, i.e.~they have transition functions with values in $\sfO(n,n;\IZ)$. These can be produced by T-dualities involving $F^1$-backgrounds. The second step is to go even beyond those, which leads to the \emph{R-spaces} obtained e.g.~by T-dualizing $F^0$-backgrounds. These do not even come with a local geometric description, and interpretations in terms of non-associative geometry~\cite{Bouwknegt:2004ap}  
  and double field theory (cf.~\cite{Plauschinn:2018wbo,Szabo:2018hhh}) have been put forward. In~\cite{Kim:2022opr}, a description in terms of higher geometry has been proposed which we will sketch in the following, focusing on the big picture.
  
  \subsection{T-duality with T-folds}
  
  The generalization to T-folds is particularly straightforward: we simply remove the restriction to geometric principal $\scTD_n$-bundles, and consider general such bundles, allowing T-duality transformations to glue together the scalar fields. 
  
  We note that the description of half-geometric topological T-dualities presented in~\cite{Nikolaus:2018qop} is subsumed in our proposal. In a half-geometric T-duality, one of the T-dual backgrounds is still purely geometric, while our description allows for both backgrounds to be T-duals. Moreover, for trivial scalar fields, we recover our above description of geometric T-dualities in terms of principal $\underline{\sfTD}_n$-bundles.
  
  As a concrete example, let us again consider the popular example of the nilmanifold we encountered in section~\ref{ssec:full_T_duality}. Here, we wish to perform a T-duality along both the fiber and the $y$-direction. This is well-known to produce a T-fold. In our proposal, this amounts to a principal $\scTD_n$-bundle $\scP_\text{C}$ over the circle $S^1$. Principal torus bundles over the circle are trivial; similarly, most of the cocycle data trivializes. Eventually, the bundle $\scP_\text{C}$ reduces to a functor from the \v{C}ech groupoid to the scalar groupoid~\eqref{eq:scalar_groupoid}. Explicitly, we have
  \begin{equation}
    \begin{gathered}
      q:C^\infty(Y,Q_2)\eand g:C^\infty(Y^{[2]},\sfGO(n,n;\IZ))~,
      \\
      g_{ij}\acton q_j=q_i\eand g_{ij}g_{jk}=g_{ik}~.
    \end{gathered}
  \end{equation}
  The $q_i$ are the local scalar fields, and, as expected for a T-fold, this local data is glued together with a T-duality transformation $g_{ij}$. The target space of the scalar fields $Q_n\cong \IR^4$ arises from the dimensional reduction of the metric and the Kalb--Ramond field in two directions, with the components $g_{yy},g_{yz},g_{zz}$, and $B_{yz}$. In order to manifest the action of $\sfGO(2,2;\IZ)$, we again arrange these moduli in the form of a generalized metric. 
  In the example at hand, we can write 
  \begin{equation}
    \caH_H(x)=\begin{pmatrix}
      1+B_{yz}^2 & 0 & 0 & B_{yz}
      \\
      0 & 1+B_{yz}^2 & -B_{yz} & 0
      \\
      0 & -B_{yz} & 1 & 0
      \\
      B_{yz} & 0 & 0 & 1
    \end{pmatrix}~,~~B_{yz}=\ell x~,
  \end{equation}
  and $\sfGO(2,2;\IZ)$ acts by conjugation. Recall that the T-duality transformations given by the flip automorphism~\eqref{eq:flip_automorphism} are themselves elements of $\sfGO(2,2;\IZ)$. We can now use these to act on the above generalized metric $\caH_H(x)$ to obtain two further generalized metrics:
  \begin{equation}
    \begin{aligned}
      \caH_f(x)&=
      \begin{pmatrix}
        1 & -B_{yz}^2 & 0 & 0 
        \\
        -B_{yz} & 1+B_{yz}^2 & 0 & 0
        \\
        0 & 0 & 1+B_{yz}^2 & B_{yz}
        \\
        0 & 0 & B_{yz} & 1
      \end{pmatrix}~,~~~
      \caH_Q(x)&=
      \begin{pmatrix}
        1 & 0 & 0 & -B_{yz}
        \\
        0 & 1 & B_{yz} & 0
        \\
        0 & B_{yz} & 1+B_{yz}^2 & 0
        \\
        -B_{yz} & 0 & 0 & 1+B_{yz}^2
      \end{pmatrix}~.
    \end{aligned}
  \end{equation}
  Comparing these generalized metrics to the general form~\eqref{eq:generalized_metric} allows us to interpret these backgrounds as follows. The generalized metric $\caH_H$ describes a 3-torus $T^3$ since $g=\unit_2$, carrying a gerbe with non-trivial $B$-field $B_{yz}=\ell x$. The generalized metric $\caH_f$ has trivial $B$-field, but a non-trivial metric, indicating a non-trivial principal circle bundle over $T^2$, but the corresponding $B$-field is trivial. This is the case of the nilmanifold $N_\ell$. The last generalized metric $\caH_Q$ comes with a metric which is not globally defined but glued together by $\sfGO(2,2;\IZ)$-transformations. This describes the T-fold derived in the literature by T-dualizing a 3-torus carrying $H$-flux in two directions.
  
  \subsection{T-duality with R-spaces}
  
  There remains one more step, namely T-dualities potentially involving $F^0$-backgrounds. For such backgrounds, the 3-form curvature $H$ aligns completely with the T-duality directions. Our proposal for this description is physically speculative, but mathematically quite elegant. 
  
  In this case, the dimensional reduction that reduces the Kalb--Ramond field $B$ to $1$-forms and scalar fields produces additional $-1$-forms, which are clearly non-sensical. Note, however, that the $2$-, $1$-, and $0$-forms all come with corresponding curvatures, which are global exact differential forms of one degree higher. Analogously, we can replace the non-existing $-1$-forms with a global ``curvature'' $0$-form. It remains to clarify in which space these $0$-forms should take values. The crucial input here is the observation that non-geometric fluxes essentially define (at least parts of) the embedding tensor~\cite{Aldazabal:2011nj,Geissbuhler:2011mx,Grana:2012rr} in the tensor hierarchy of gauged supergravity. 
  
  Without going into further details, we note that this tensor hierarchy produces the kinematic data of an adjusted higher gauge theory that then becomes part of a supergravity theory. Moreover, it provides the best understood and very rich class of examples of adjusted connections~\cite{Borsten:2021ljb}. A detailed analysis of the constraints that one would impose on the space of embedding tensors $\bar R_n$ then shows that it carries precisely the expected symmetry structure~\cite{Kim:2022opr}.
  
  It remains to extend the Lie 2-groupoid by the space $\bar R_n$, for which we will need to introduce augmented Lie 2-groupoids. Recall that simplicial sets can be \emph{augmented} by introducing additional $-1$-simplices. A very natural example of this augmentation is the nerve of the \v{C}ech groupoid of a surjective submersion $Y\rightarrow X$, which comes with the natural augmentation by the manifold $X$:
  \begin{equation}\label{eq:augmented_Cech_nerve}
    \check\scC_\text{aug}(Y\rightarrow X)\ \coloneqq\ \left(\begin{tikzcd}
      \ldots \arrow[r,shift left,shift left,shift left] 
      \arrow[r,shift right,shift right,shift right]
      \arrow[r,shift left] 
      \arrow[r,shift right] & Y^{[3]} \arrow[r,shift left,shift left] 
      \arrow[r,shift right,shift right]
      \arrow[r] &
      Y^{[2]}\arrow[r,shift left] 
      \arrow[r,shift right] & Y \arrow[r] & X
    \end{tikzcd}\right)~.
  \end{equation}
  Similarly, we augment the 2-groupoid $\scTD_n$ by $\bar R_n$, where we have an additional action of $\sfGO(n,n;\IZ)$, and hence of $\scGO(n,n;\IZ)$, on $\bar R_n$. This action is induced from the interpretation of this space as the space of embedding tensors. The result is the augmented 2-groupoid $\scTD_n^\text{aug}$.
  
  Our proposal is then that a non-geometric T-duality is captured by a principal $\scTD_n^\text{aug}$-bundle with adjusted connection, which is essentially straightforward to define given the above data, cf.~\cite{Kim:2022opr}. We note that such a bundle contains indeed a global map from the manifold $X$ into the space of embedding tensors. We identify these with the $R$-fluxes in the physics literature.
  
  \section{Conclusions}
  
  We have reviewed a proposal for a description of geometric and non-geometric T-dualities in terms of higher groupoid bundles with connection, which passes a number of non-trivial consistency checks and possesses many expected properties.
  
  Our proposal for the most general T-dualities in terms of principal $\scTD_n^\text{aug}$-bundle (with adjusted connections) can capture $R$-fluxes as global scalar fields and naturally specializes to T-dualities with locally geometric backgrounds given in terms of principal $\scTD_n$-bundles. This specialization incorporates the additional scalar fields expected from the Kaluza--Klein reduction underlying T-duality, and we reproduce the sequence of T-dualities taking a 3-torus $T^3$ with $H$-flux to a T-fold. Moreover, the principal $\scTD_n$-bundles contain in particular the topological description of half-geometric T-dualities of~\cite{Nikolaus:2018qop}. Our proposal further specializes to the purely geometric T-dualities, which are described by geometric principal $\scTD_n$-bundles. In this case, we reproduce the well-known T-dualities between nilmanifolds carrying $H$-flux. Also, as shown in~\cite{Waldorf:2022tib}, our picture locally yields the Buscher rules. Finally, we note that in the original paper~\cite{Kim:2022opr}, the geometric T-dualities were extended to affine torus bundles in the sense of~\cite{Baraglia:1105.0290}; we have dropped these from our discussion here for pedagogical reasons.
  
  Our proposal clearly needs much further study, mostly in the non-geometric cases. In particular, one should verify the consistency with the various physical expectations found in the literature. An interesting point is that the constructions suggest that non-commutative and non-associative geometries, at least those associated with non-geometric backgrounds arising from T-dualities, may be resolved into ordinary but higher geometries. 
  
  In the future, we plan to study various extensions; in particular, non-abelian and Poisson Lie T-duality come to mind. The long-term goal is certainly a description of U-duality along similar lines.   
  
  \section*{Acknowledgments}
  
  We would like to thank Konrad Waldorf for helpful discussions. The work of HK and CS was partially supported by the Leverhulme Research Project Grant RPG-2018-329 ``The Mathematics of M5-Branes.''
  
  \bibliographystyle{latexeu}
  \bibliography{bigone}

  %
  
\end{document}